\documentclass[%
 reprint,
 amsmath,amssymb,
 aps,
]{revtex4-1}

\usepackage[pdftex]{graphicx}
\usepackage{dcolumn}
\usepackage{bm}
\usepackage{float}

\begin{document}


\title{CMS Measurements of the
Higgs-like Boson \\In the
Two Photon Decay Channel}%

\author{Christopher Palmer}
 \affiliation{Physics Department, University of California - San Diego}
 \collaboration{CMS Collaboration}

\date{\today}

\begin{abstract}
CMS reports on the recently updated, preliminary results with the full datasets of 
2011 and 2012 in the analysis of the Higgs-like Boson at 125 GeV.  Utilizing 5.1$fb^{-1}$
of 7 TeV data and 19.6$fb^{-1}$ of 8 TeV data, a signal strength of
$0.78^{+0.28}_{-0.26}$ times the Standard Model (SM) expectations with a mass of 
$125.4\pm0.8$ GeV is observed.  The significance of this resonance with respect to the 
background only prediction is $3.2\sigma$.  The cut-based cross-check analysis observes
signal strength of $1.11^{+0.32}_{-0.30}$ times the SM expectations with a significance 
of $3.9\sigma$ at 124.5 GeV.
\end{abstract}

\maketitle


\section{\label{sec:level1} Analysis Motivation}

On the fourth of July, 2012, the discovery of a new particle was announced
\cite{Discovery}.  A previously unobserved particle was discovered in two channels near 
125 GeV, while searching for this Higgs boson, a particle central to the 
SM\cite{SM1,SM2,Higgs1,Higgs2}. One of the channels in which this particle is observed is
in a two photon resonance.  CMS has improved, expanded, and updated the analysis of the 
two photon channel, which is one of the two channels in which the discovery was made.  
This update contains CMS's first preliminary results on the entire datasets from 2011 and
2012.

\section{\label{sec:level2} Analysis Description}

The decay signature sought in this analysis is a small, narrow resonance on top of a 
large, monotonically decreasing background.  This analysis has two basic features:  
background estimates from fits to the data and, categorization of data and monte carlo 
signal.

Since the data is dominated by background, it is possible to fit 3rd to 5th order 
polynomials to the data for background  estimates.  A thorough bias study was performed 
to ensure that potential bias, calculated using numerous, varied underlying shape 
assumptions, would be less than 20\% of the uncertainty from the polynomial fits.

The other general feature of the analysis is categorization of events.  Events with
different $S/\sqrt{B}$ and resolution are best used when used with similar events. In 
order to optimally categorize a Boost Decision Tree (BDT) is used. The BDT takes several 
inputs and gives a single output, which is used to categorize events.  

In addition, there are several final state signatures which are indicative of the 
production mechanism of the boson.  These events feature extra high transverse momentum 
objects which can be tagged.  An additional category, which has higher than normal 
$S/\sqrt{B}$, is created for each tagged class.

\subsection{\label{sec:level2} Vertex Selection and Vertex Probability}

On average, events from the 2012 dataset contain 20 interactions. The challenge for this 
analysis is to determine the vertex accurately in order to improve the mass resolution of 
the reconstructed di-photon.  Since there are no tracks associated to photons, except in 
the case of reconstructed conversions, obtaining the correct vertex is not 
straightforward, particularly for di-photons with low transverse momentum.

A BDT takes information from conversions, recoil between vertices' tracks and the 
di-photon system, and the sum of vertices' track transverse momentum to sort the 
reconstructed vertices.  The one with the highest score is selected to be used in the 
mass calculation.

A second BDT is trained after selecting the vertex to determine the probability of 
having chosen the vertex correctly. This BDT uses information from the chosen vertex, as 
well as the second and third ranked vertices to determine a probability of accurate 
selection.  This probability is measured from efficiency in simulation and is fully 
validated using Z boson events decaying to two muons where the muon tracks have been 
removed.

\subsection{\label{sec:level2} Photon Identification MVA (Multi-Variate Analysis)}

Distinguishing real photons from fake reconstructed photons, which are mainly jets, is a 
very important aspect of this analysis.  There are two main types of variables
which are used for this purpose:  shower topology and isolation.  There are numerous 
shower topology variables, in which the profile of fake photons differs that of real 
photons.  Isolation variables are powerful because real photons typically do not have 
very many particles in the immediate vicinity of the main energy deposit of the photon.

These variables, along with a variable correlated with number of simultaneously recorded
interactions and the detector location of the photons, are given as input to a BDT.  The
output, referred to as Photon ID MVA, is validated and shape-corrected using Z boson 
decay to 2 electrons where the electrons are treated as photons.  In 
figure the distributions of the Photon ID MVA for data and monte carlo 
for Z boson events are shown.



\subsection{\label{sec:level2} Di-Photon MVA}

This BDT, referred to as Di-Photon MVA, takes several inputs: di-photon kinematics, per 
event resolution estimates for right and wrong vertex assumptions, the vertex probability,
the Photon ID MVA scores of both selected photons, and the detector location of the 
photons.  The Di-Photon MVA is used to categorize events into four optimized bins.

\begin{figure}[H]
\includegraphics[width=85mm]{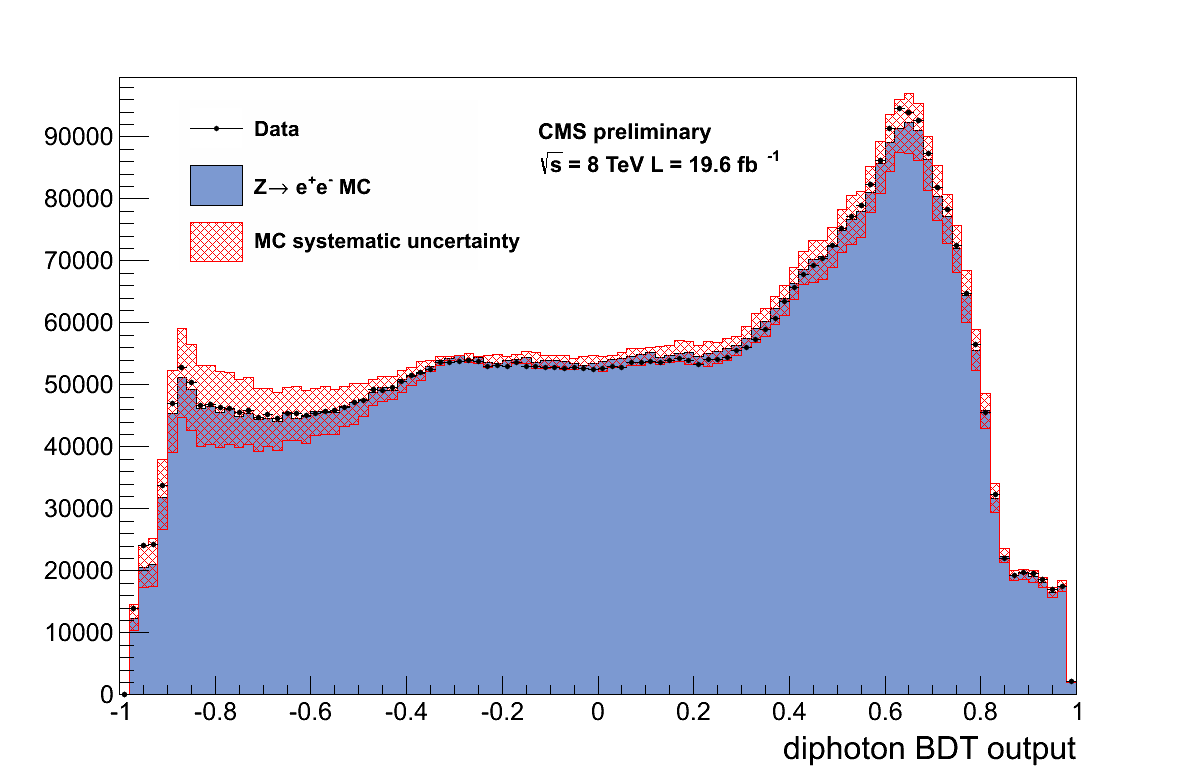}
\caption{\label{fig:diphomva}
Comparison of the diphoton MVA output for Z$\rightarrow$ee events in data and Monte Carlo where the electrons are reconstructed as photons, for events passing the pre-selection of the mass-fit MVA analysis (with the electron veto condition inverted). Corrections have been applied to the MVA inputs for MC.}
\end{figure}

The Di-Photon MVA is validated using Z boson to two electron events where the electrons 
are treated like photons. In figure~\ref{fig:diphomva} the Di-Photon MVA is shown for the 
Drell-Yan samples in data and in simulation.  Agreement within systematic error is shown.

\subsection{\label{sec:level2} Exclusive Channels}

The dominant production mechanism of SM higgs is gluon fusion.  However, The SM production
of higgs via Vector Boson Fusion (VBF) and Higgsstrahlung (VH) is non-negligible.  Higgs 
produced via these mechanisms contain high transverse momentum objects, which can be 
tagged.  In this analysis several tagged categories have been made to utilize these 
additional objects.

The di-jet tags and the most powerful is of these tags, which are intended to tag the two 
forward jets from VBF production.  A BDT is used to categorize the events with two jets 
and there are two categories.  This BDT uses mostly jet kinematics, some
di-jet plus di-photon kinematics and some photon kinematics. These two categories account 
for nearly 20\% of the sensitivity of the analysis.


In addition to the di-jet tagged categories, there are three other categories, which 
intend to tag the leptonic decays of vector bosons, which produce higgs via associated
production (Higgsstrahlung).  The three tags of muons, electrons and MET greatly reduce 
the uncertainty of the measured coupling of the higgs to the vector bosons.

\begin{figure}[H]
\includegraphics[width=85mm]{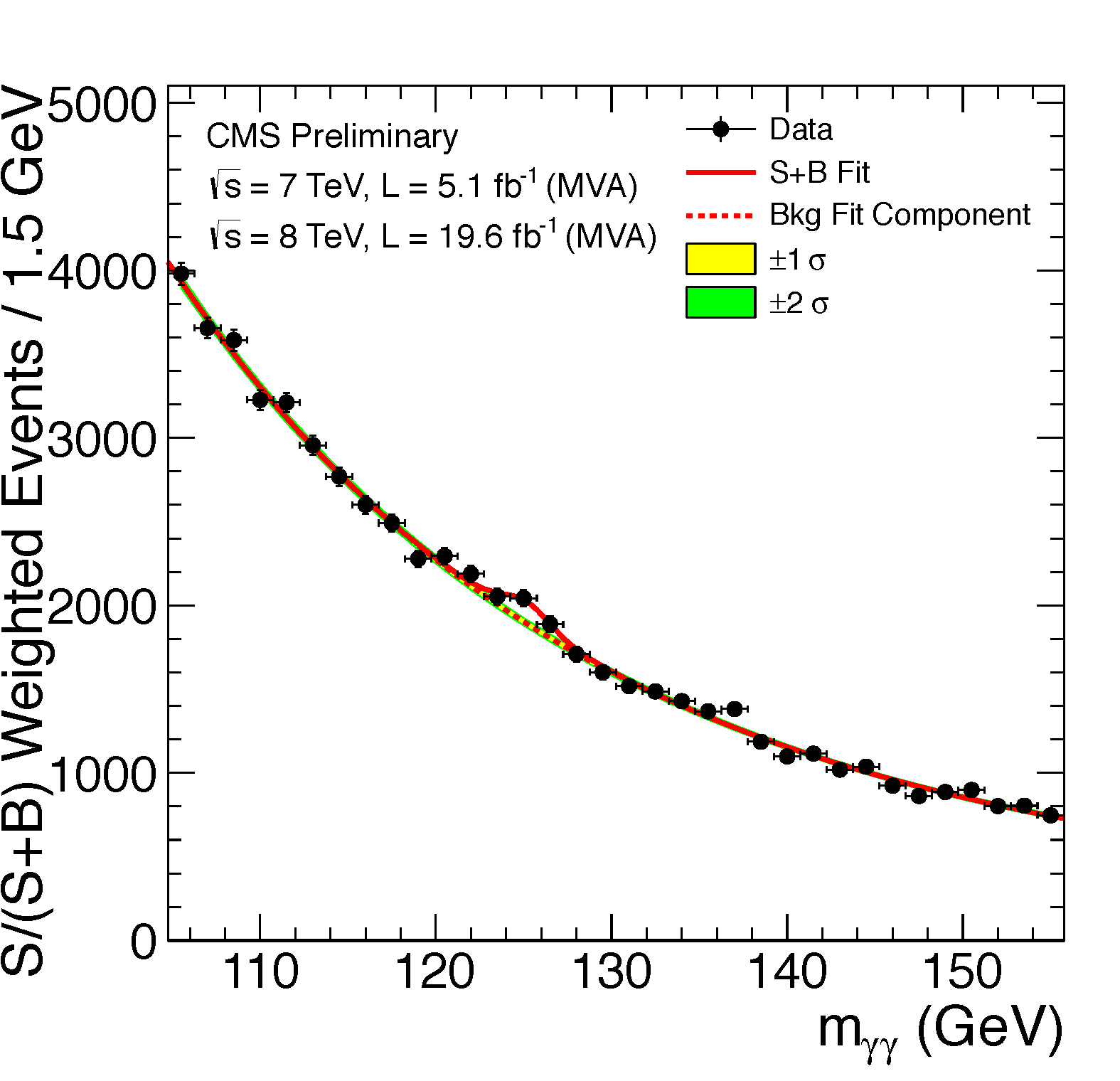}
\includegraphics[width=85mm]{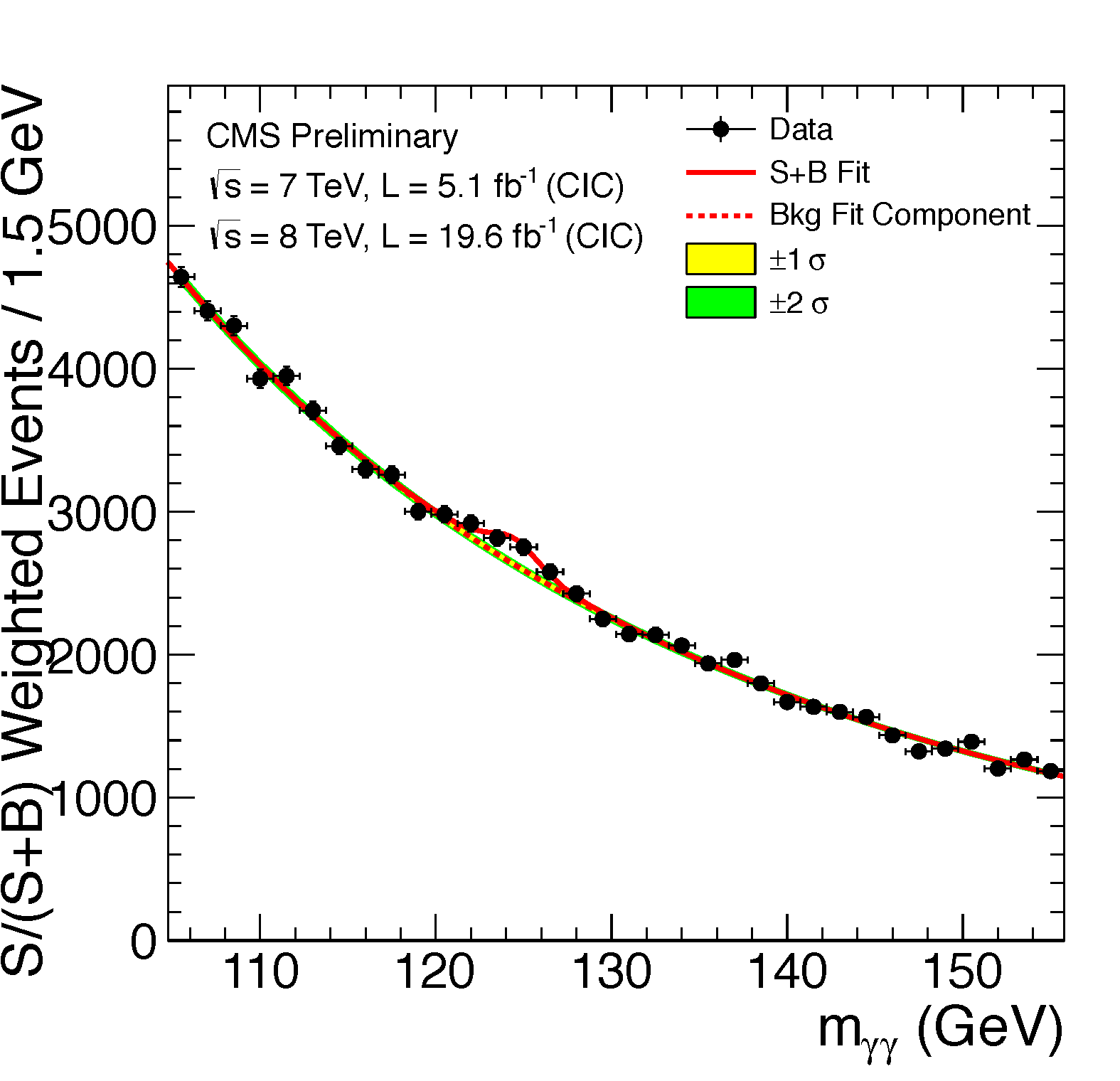}
\caption{\label{fig:data}
The data and background plus signal fits for the categories in the main (above) and 
cut-based (below) analyses have been weighted by S/(S+B) and summed.  The normalization 
is such that the signal yield is same as the best fit. }
\end{figure}

\subsection{\label{sec:level2} Cut-based Analysis}

In addition to the full analysis outlined already, a simple, robust, cut-based analysis is
performed in parallel.  It utilizes the main features of the main analysis, which are 
categorization and polynomial background fits to the data, but it uses cut-based photon
selection and categorization based on the detector location of the photons and their 
shower shapes.

There are analogous exclusive tags in the cut-based analysis.  The di-jet tag categories 
use a cut-based approach with the same inputs as the di-jet MVA in the main analysis.  The
muon, electron, and MET tags are nearly identical except for the cut-based selection of 
the photons.

\subsection{\label{sec:level2} 2011 Data Analysis}

The 2011 dataset is analyzed separately and is identical to that in the discovery paper 
\cite{Discovery}. For the untagged categories, the same strategy is used with a BDT 
trained on 7 TeV monte carlo.  The only exclusive channel is a single, cut-based, di-jet 
tag.  The cut-based analysis also uses an analogous strategy and uses the same cut-based, 
di-jet tag as the MVA analysis.

\section{\label{sec:level2} Results - 2011 and 2012 Data}

\begin{figure}[H]
\includegraphics[width=85mm]{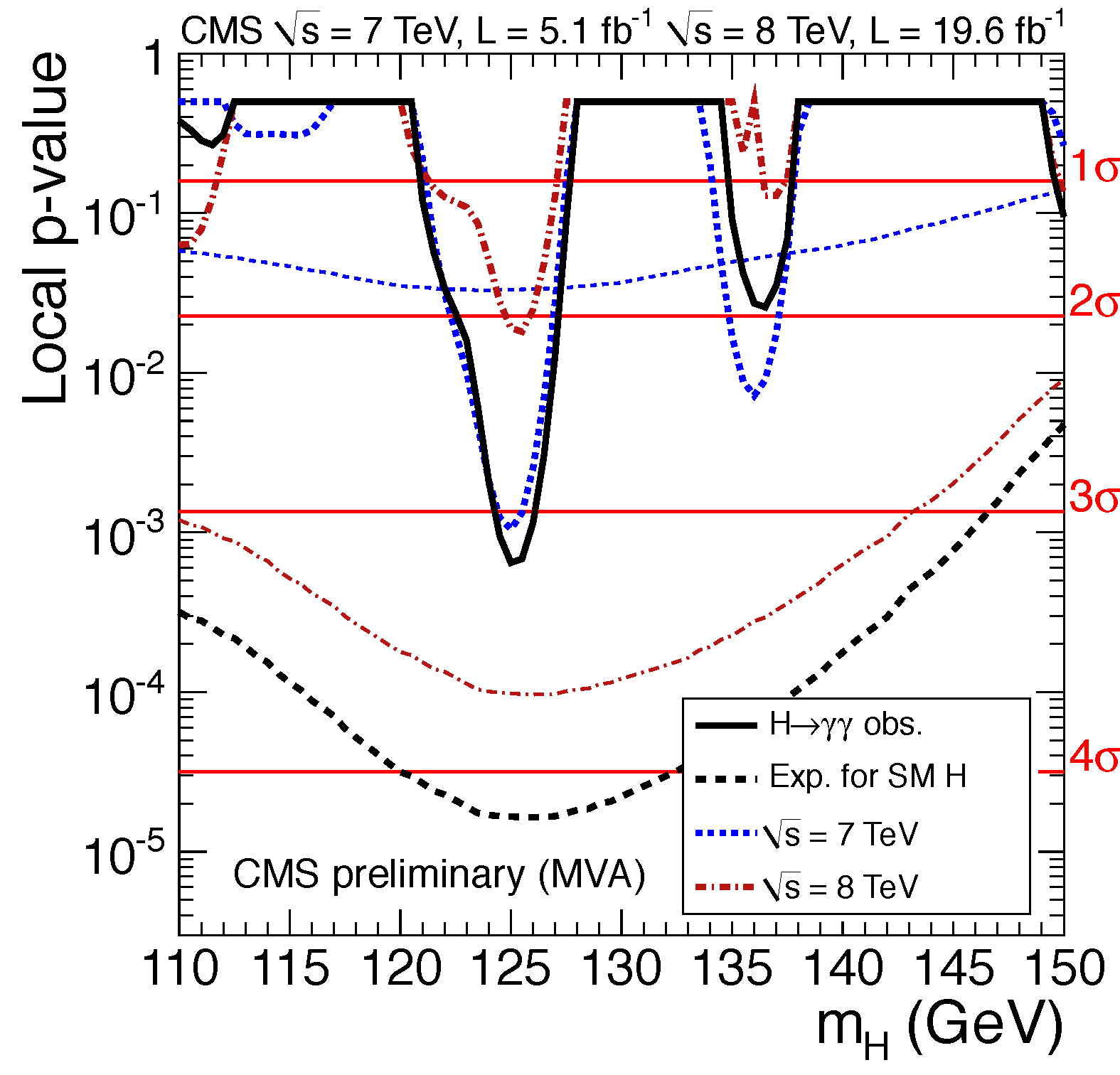}
\includegraphics[width=85mm]{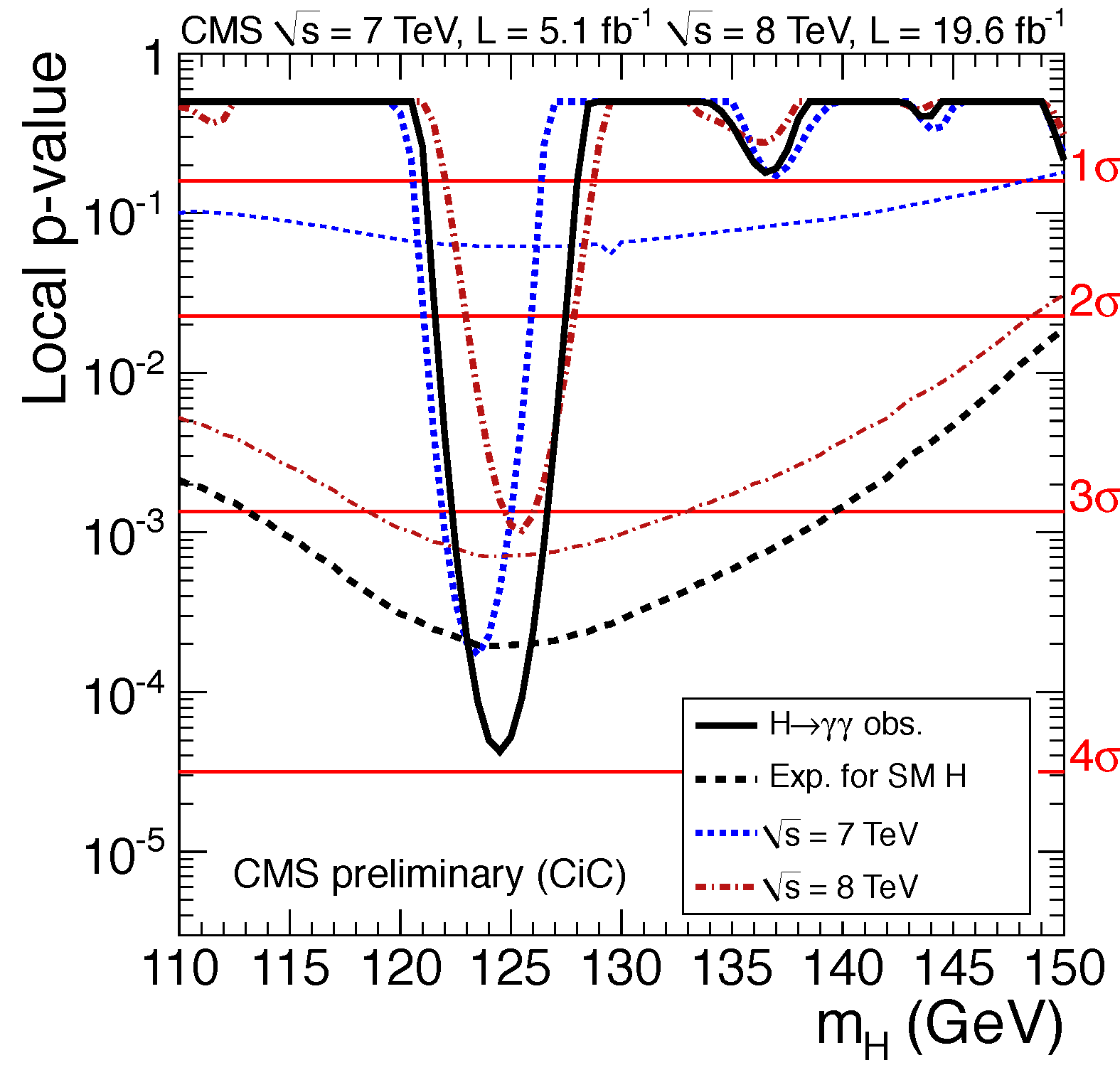}
\caption{\label{fig:pvalues}
Observed local p-values as a function of mH obtained with the mass fit MVA analysis for the 7 and 8 TeV datasets.  The main analysis is above and the cut-based is below.}
\end{figure}

The S/(S+B) weighted distributions of data in the MVA and cut-based analyses are shown in
figure~\ref{fig:data}.  The expected and observed significance is in 
figure~\ref{fig:pvalues} for the two analyses. At 125 GeV, the expected significance is 
$4.2\sigma$ and the observed is $3.2\sigma$ in the main analysis.   At 124.5 GeV, the 
expected significance is $3.5\sigma$ and the observed is $3.9\sigma$ in the cut-based 
analysis.  

\begin{figure}[H]
\includegraphics[width=85mm]{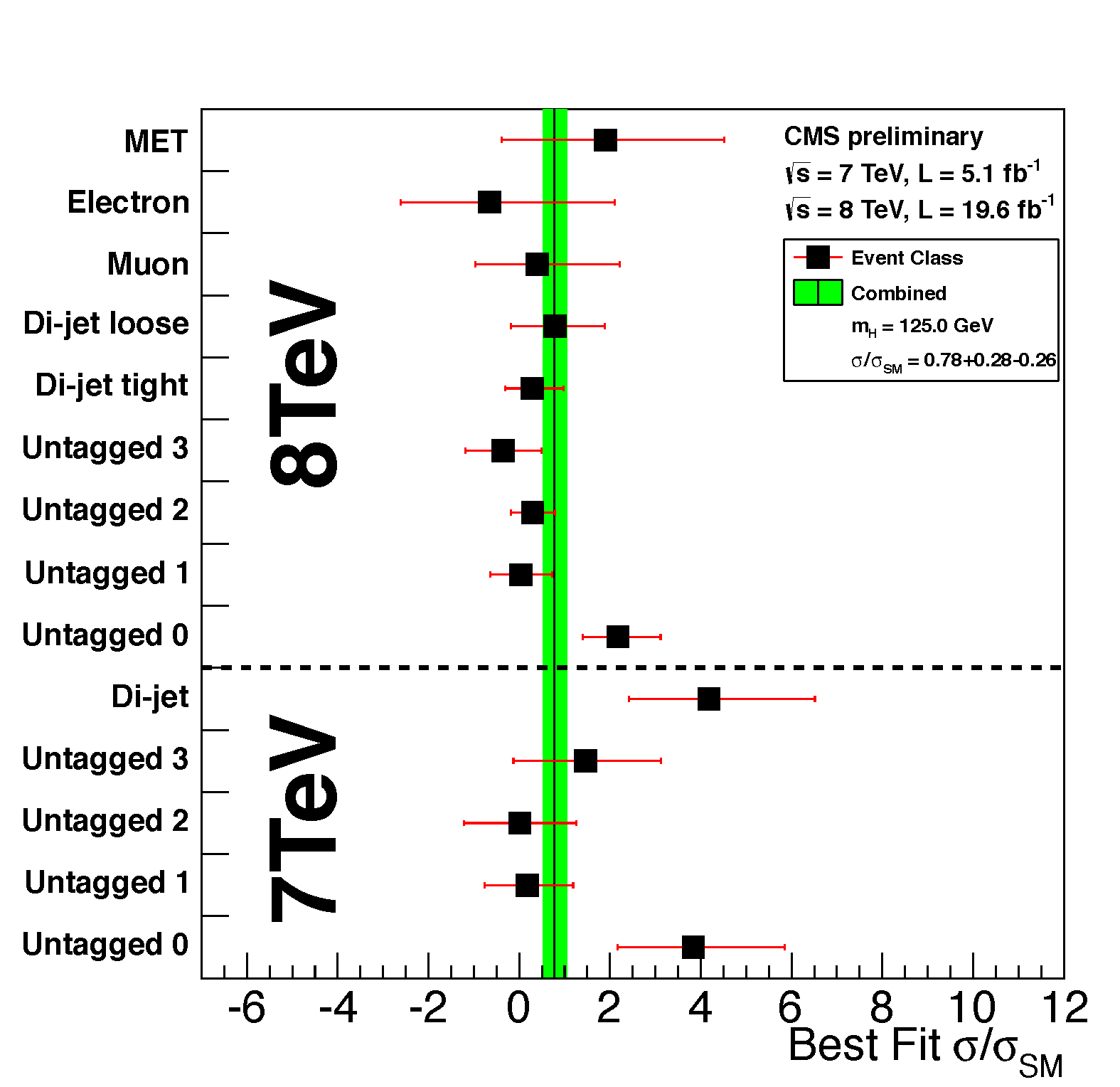}
\caption{\label{fig:chancompat}
The best fit signal strength, ($\sigma/\sigma_{SM}$), obtained in the mass-fit-MVA analysis for the combined fit to the five classes used for the 7 TeV data set and the nine classes used for the 8 TeV data set (vertical line) and for the individual contributing classes (points) for the hypothesis of a SM Higgs boson mass of 125.0 GeV. }
\end{figure}

In figure~\ref{fig:chancompat} is the channel compatibility of each of the 
categories and the combination is in green.  In the 2011 data one can see the large 
excess in untagged 0 and di-jet tag. In the 2012 data a large excess is visible in 
untagged 0 as well.

\begin{figure}[H]
\includegraphics[width=85mm]{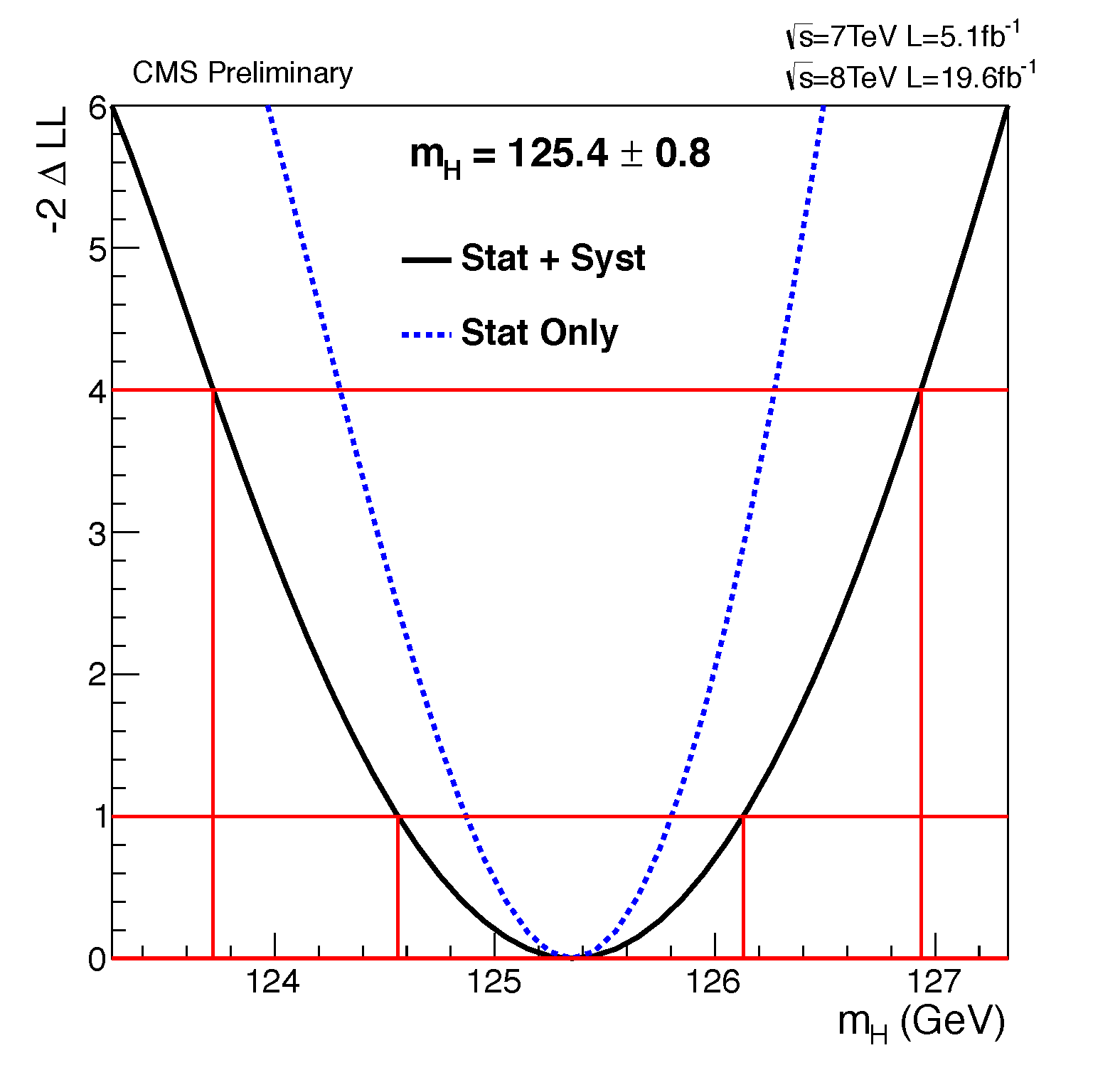}
\caption{\label{fig:mhNLL}
The 1D test statistic –2 ln Q vs Higgs boson mass hypothesis, mH. The solid line is obtained with all nuisance parameters profiled and so includes both statistical and systematic uncertainties. The dashed line is obtained with all nuisance parameters fixed to their best-fit values and thus includes only statistical uncertainties. The crossings with the thick (thin) horizontal lines define the 68\% (95\%) CL interval for the measured mass. To reduce model-dependency, two production cross-section scaling factors ($\mu_{ggH+t\bar{t}H}$ and $\mu_{VBF+VH}$) are introduced as nuisance parameters and profiled in this measurement.}
\end{figure}

An important property of the observed boson is its mass.  In this decay channel 
there is very good mass resolution.  The two main systematics that limit this measurement
are the differences between electrons and photons, and the uncertainty in energy scale
which is extrapolated from the mass of the Z boson to the mass of the observation.  The 
combined systematic from these two has been estimated to be 0.47\%. The mass of the 
higgs-like particle has been measured at be $125.4\pm0.5(stat.)\pm0.6(syst.)$ GeV and 
the measurement is shown in figure~\ref{fig:mhNLL}.

\begin{figure}[H]
\includegraphics[width=85mm]{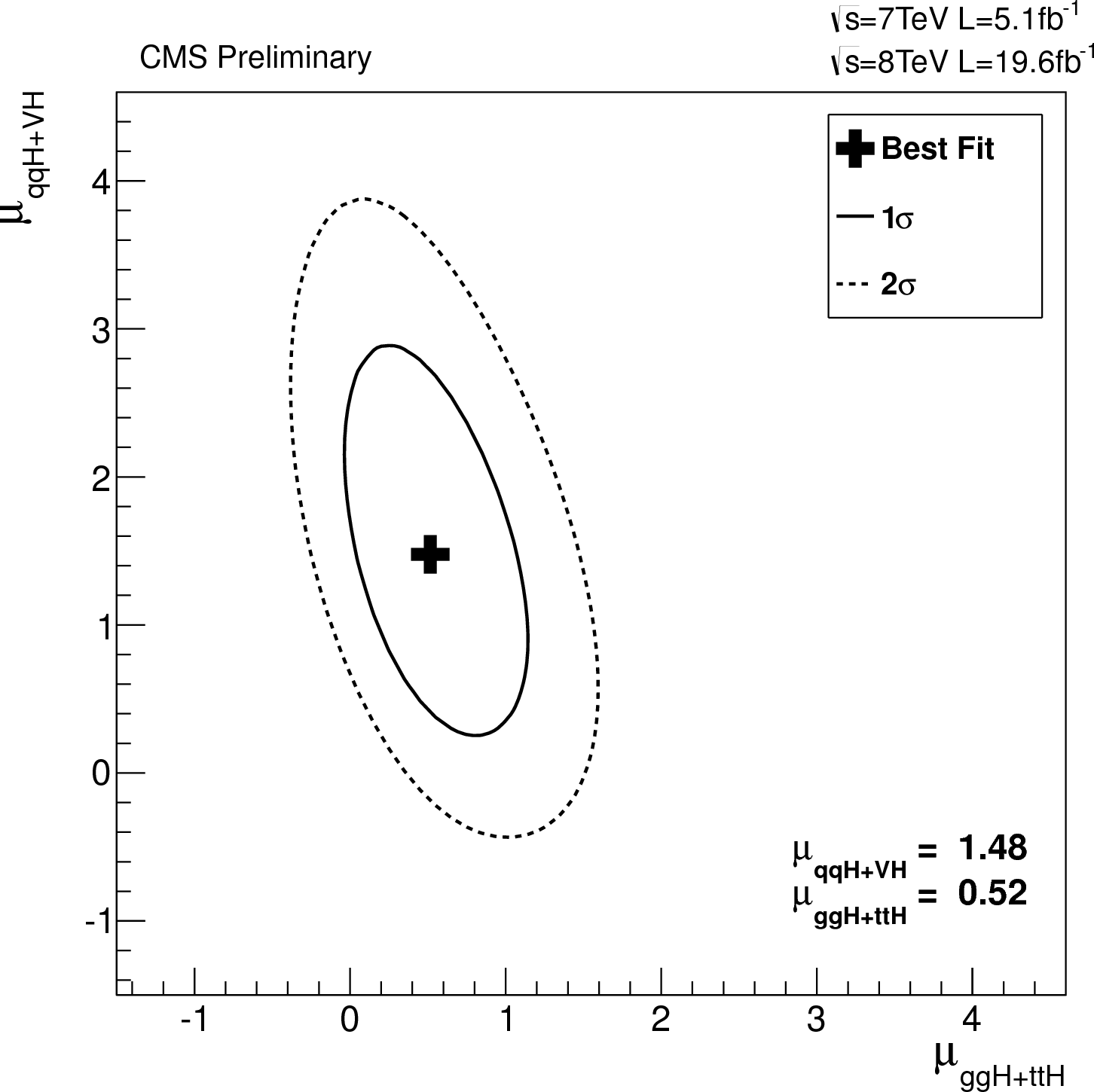}
\caption{\label{fig:rvrf}
The 68\% CL (solid line) and 95\% CL (dashed line) intervals for signal strength in the gluon-gluon-fusion-plus-ttH and in VBF-plus-VH production mechanisms: $\mu_{gg+ttH}$ and $\mu_{VBF+VH}$, respectively.}
\end{figure}

Finally, a two dimensional scan of the signal strength was performed.  One dimension is 
the signal strength of gluon fusion production plus $t\bar{t}$H production, and the other 
is the VBF plus VH production.  In figure~\ref{fig:rvrf} the scan shows best fit values 
of $(\mu_{ggH+t\bar{t}H}=0.52,\mu_{VBF+VH}=1.48)$ with the SM prediction being 
$(\mu_{ggH+t\bar{t}H}=1.0,\mu_{VBF+VH}=1.0)$ which is within $1\sigma$ of the best fit. 
Thus, the observed boson is compatible with a SM higgs boson.

\section{\label{sec:leve1} Conclusions}

The results of the search for a Standard Model Higgs Boson decaying into two photons have
been updated using data obtained from $5.1fb^{-1}$ and $19.6fb^{-1}$ of pp collisions at 
$\sqrt{s} = 7$TeV and at $\sqrt{s} = 8$TeV, respectively. The selected events have been 
subdivided into classes according to indicators of mass resolution and predicted 
signal-to-background ratio, and the results of a search in each class have been combined.

An excess of events above the expected SM background has been observed for a Higgs
boson mass hypothesis of 125 GeV, where the expected limit is 0.48 times the SM
expectation. The local significance of this excess is 3.2$\sigma$. This result 
constitutes further evidence for the existence of a new massive state that decays into 
two photons. The mass of the observed boson is measured to be 
$125.4\pm0.5(stat.)\pm0.6(syst.)$ GeV. For a Higgs boson mass hypothesis of 125.0 GeV, 
the best fit signal strength is $0.78^{+0.28}_{-0.26}$ times the SM Higgs boson 
cross-section.

\appendix





\end{document}